# High Pressure X-ray Diffraction Study of $MgMn_2O_4$ Tetragonal Spinel

Lorenzo Malavasi[a,*], Cristina Tealdi[a], Monica Amboage[b], M. Cristina Mozzati[c], and Giorgio Flor[a]

[a]*Dipartimento di Chimica Fisica "M. Rolla", INSTM, IENI/CNR Unità di Pavia, Università di Pavia, V.le Taramelli 16, 27100 Pavia, Italy*

[b]*European Synchrotron Radiation Facility (ESRF), 6, rue Jules Horowitz, BP 220, Grenoble, FRANCE*

[c]*INFM Unità di Pavia - Dipartimento di Fisica "A. Volta", Università di Pavia, Via Bassi 6, 27100 Pavia, Italy*

**Abstract**

The phase stability of the $MgMn_2O_4$ spinel has been studied by means of high-pressure X-ray diffraction for pressures up to 30 GPa. Two samples with different inversion degrees have been considered. Both spinels undergo a phase transition towards an orthorhombic structure ($CaMn_2O_4$-type). For the more inverted sample the transition pressure is at least 1 GPa lower with respect to that of the less inverted spinel. Also the volume contraction, relative compressibility and density trends are different for the two samples. These variations have been explained according to differences in the cation distribution. and electronic properties of the samples.

# Introduction

Spinel oxides are interesting compounds for both the basic and applicative research. For example, one of the most studied, the cubic $LiMn_2O_4$, is a promising material as cathode for Lithium batteries. Other spinel manganites with divalent cations on the tetrahedral site, such as Mn, Zn, Cd, Mg, display a tetragonal structure as a consequence of a strong Jahn-Teller (J-T) effect due to the presence of Mn(III) ions that fill the octahedral sites of the oxygen ion closed packed arrangement.

The literature regarding the high-pressure (HP) structural properties of spinel manganites is relatively scarce if compared to other materials. However, a detailed and thorough study of HP-XRD on the $Mn_3O_4$ spinel up to 38 GPa [1] and a more qualitative work on $ZnMn_2O_4$ [2] for pressure till 52 GPa are present. From these works, it turned out that both $Mn_3O_4$ and $ZnMn_2O_4$ undergo a phase transition by increasing pressure but with different HP-phases and transition pressures ($P_T$): the first spinel ($Mn_3O_4$) transforms into an orthorhombic marokite-like phase around 10 GPa while the latter changes into a tetragonal primitive cell with a significant reduction of the $c/a$ parameter around 23 GPa.

Our attention was focused on the $Mg_{1-x}Mn_{2+x}O_4$ system, for $0 \leq x \leq 1$, which has been scarcely studied in the previous literature [3-5]. This spinel, opposite to $ZnMn_2O_4$ and $CdMn_2O_4$, has a greater tendency to be inverted which means that part of magnesium ions can be found in the octahedral sites already at room temperature. The inversion degree is usually expressed by a parameter, $m$, and following the spinel formula is written as: $(A_{1-m}B_m)_{tet}[A_mB_{2-m}]_{oct}O_4$.

Object of this work is the study of the high-pressure phase stability of the $MgMn_2O_4$ spinel considering also the role of cation distribution in the lattice. For this purpose we studied a first sample with $m \approx 0.2$ and a second sample with $m \approx 0.4$.

# Experimental

Synthesis and preparation of $MgMn_2O_4$ samples with different inversion degrees ($m \approx 0.2$ and $m \approx 0.4$) are reported in detail in Ref. 3.

The high-pressure powder diffraction experiments were carried out at the European Synchrotron Radiation Facility (ESRF) of Grenoble, France, on the ID09 beamline by employing a diamond anvil cell (DAC) with 350 μm diameter culets. The $N_2$ was used as pressure transmitting medium. Patterns were collected at intervals of about 1 GPa.

The *P* calibration was accomplished by following the fluorescence line of ruby excited by an Ar laser source. Diffraction images were collected at a wavelength of $\lambda = 0.41793$ Å.

Rietveld refinements of the XRD patterns were performed by means of FULLPROF software package [6] while equation of state analysis (EoS) was performed by means of the EOSFIT Software [7].

## Results and Discussion

Figure 1 reports the volume variation for the first sample considered, *i.e.* the one with the lower inversion degree, as calculated from the refinement of the diffraction patterns. Data up to 15.6 GPa are indexed according to the tetragonal spinel structure $I4_1/adm$ (space group 141) while above this pressure a phase transition occurs. The HP phase has been identified as an orthorhombic structure analogous to the $CaMn_2O_4$ one (s.g. *Pmab*, no. 57), similarly to what happen to the $Mn_3O_4$ compound [1].

Up to the transition pressure ($P_T$) the cell volume progressively contracts of about 7.5% At the phase transition the cell contracts of 10.2% and the bulk density increases from 4.70 to 5.24 g/cm$^3$ (+11.4%). For comparison, in the $Mn_3O_4$, at the $P_T$, which is between 10 and 12 GPa, the volume reduces of 8.6% [1].

The *V vs. P* data presented in Figure 1 were used to calculate the bulk modulus according to the Murnaghan equation of state (EoS). $K'$ was fixed to 4. For the LP-phase (Low-Pressure) the $K_0$ value was 156±0.7 GPa while for the orthorhombic HP-phase it increased to 196±4 GPa. For the $Mn_3O_4$ $K_0$ = 134±3.7 GPa for the tetragonal phase and 166.6±2.7 GPa for the marokite-like HP structure [1]. For $ZnMn_2O_4$ $K_0$ = 197±5 GPa by considering *V vs. P* data up to only 12 GPa [2]. Finally, for most of the cubic spinels which become orthorhombic at HP the differences between the LP and HP phases, in term of bulk modulus and density, are smaller with respect to the tetragonal spinels; for example, the HP-structures are generally only 2-3% denser than the cubic LP ones. This smaller difference between the HP and LP polymorphs is reasonable since the cubic structure is more compact with respect to the tetragonal one for which, in fact, the density variation between the two polymorphs is significantly larger.

Let us consider now the sample with a higher inversion degree ($m \approx 0.4$). Also for this sample a structural phase transition (*tetragonal* → *orthorhombic*) occurred around 14.4 GPa. The new phase is stable up to the highest $P$ studied, 23 GPa.

Figure 2 reports the normalized lattice parameters of this sample (full symbols) together with the ones of sample with lower inversion (empty symbols).

The comparison between the normalized lattice parameters shows that small but significant differences are present. The compressibility along the $c$ axes slightly reduces while the one along the $a$ axis increases by increasing the inversion. Let us note that in the case of a cubic structure no differences between the three directions should be present. So, this result can be directly connected to the increased inversion in the spinel and in the change of cation occupancies between the tetrahedral and octahedral structural sites.

In Figure 3 is plotted the volume variation of the sample with $m \approx 0.4$ against applied pressure. The volume contraction at the $P_T$ is around 9.5%. Again, the HP phase was indexed according to the $CaMn_2O_4$ structure. Across the transition the theoretical bulk density increases of about 10.7% from 4.694 g/cm$^3$ (14.5 GPa) to 5.187 g/cm$^3$ (16.2 GPa). The $V$ reduction is 7.6% from 0 to 14.5 GPa. The bulk modulus of this second sample was found to be 155.1±1.2 GPa. Concerning the HP phase the estimation of the $K_0$ value for this sample was not reliable mainly due to the low amount of HP data available with respect to the first sample.

# Conclusion

1. $MgMn_2O_4$ spinels, at both cation distributions, undergo a phase transition at HP towards an orthorhombic phase, as found for the $Mn_3O_4$ [1] and opposite to the $ZnMn_2O_4$ [2] spinel.
2. A correlation between the average ionic radius of the A site, $<r_A>$, for $Mn_3O_4$ and $MgMn_2O_4$ samples and the transition pressure from the tetragonal to the orthorhombic marokite-like phase exists (see Figure 4). So, for those spinels which undergo this kind of phase transition the pressure at which this happens is mainly controlled by the ionic radii and consequently by the ions distribution between the tetrahedral and octahedral sites.
3. The bulk moduli for the various spinels follow a roughly linear trend (see inset of Figure 4) thus obeying the empirical predictions of constant $K_0$-$V_0$. The tendency of tetragonal spinel manganites is to move towards $K_0$ values close to the ones found for HP polymorphs of cubic spinels, *i.e.* around 200 GPa.

Concerning the two $MgMn_2O_4$ samples we found that the more inverted sample has a lower $P_T$ with respect to the less inverted one and a smaller $V$ reduction at the $P_T$. Moreover, some differences can be observed in the relative compressibility of the crystal directions (see Figure 2).

# Figure caption

**Fig. 1.** Cell volume variation for $MgMn_2O_4$ with $m \approx 0.2$. Lines represent the EoS fit.

**Fig. 2**. Normalized lattice constants for $MgMn_2O_4$ samples with different inversion degrees: $m \approx 0.2$ (empty symbols), $m \approx 0.4$ (full symbols).

**Fig. 3**. Cell volume variation for $MgMn_2O_4$ sample with $m \approx 0.4$. Lines represent the EoS fit.

**Fig. 4**. Transition pressures for various spinel manganites (see names in the plot) *vs*. the average ionic radius of the A-site. Inset: bulk modulus values *vs*. cell volumes.

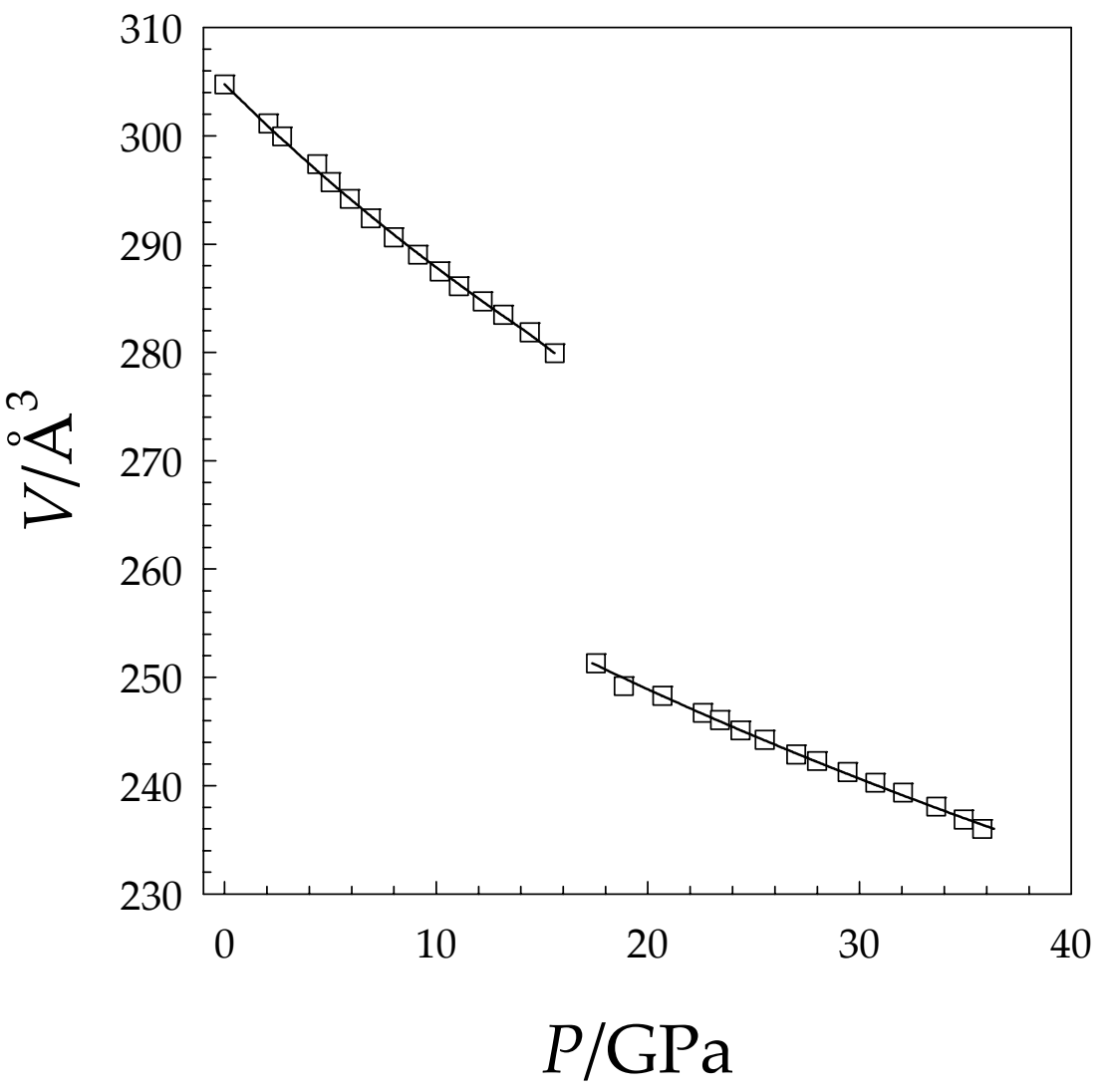

Figure 1

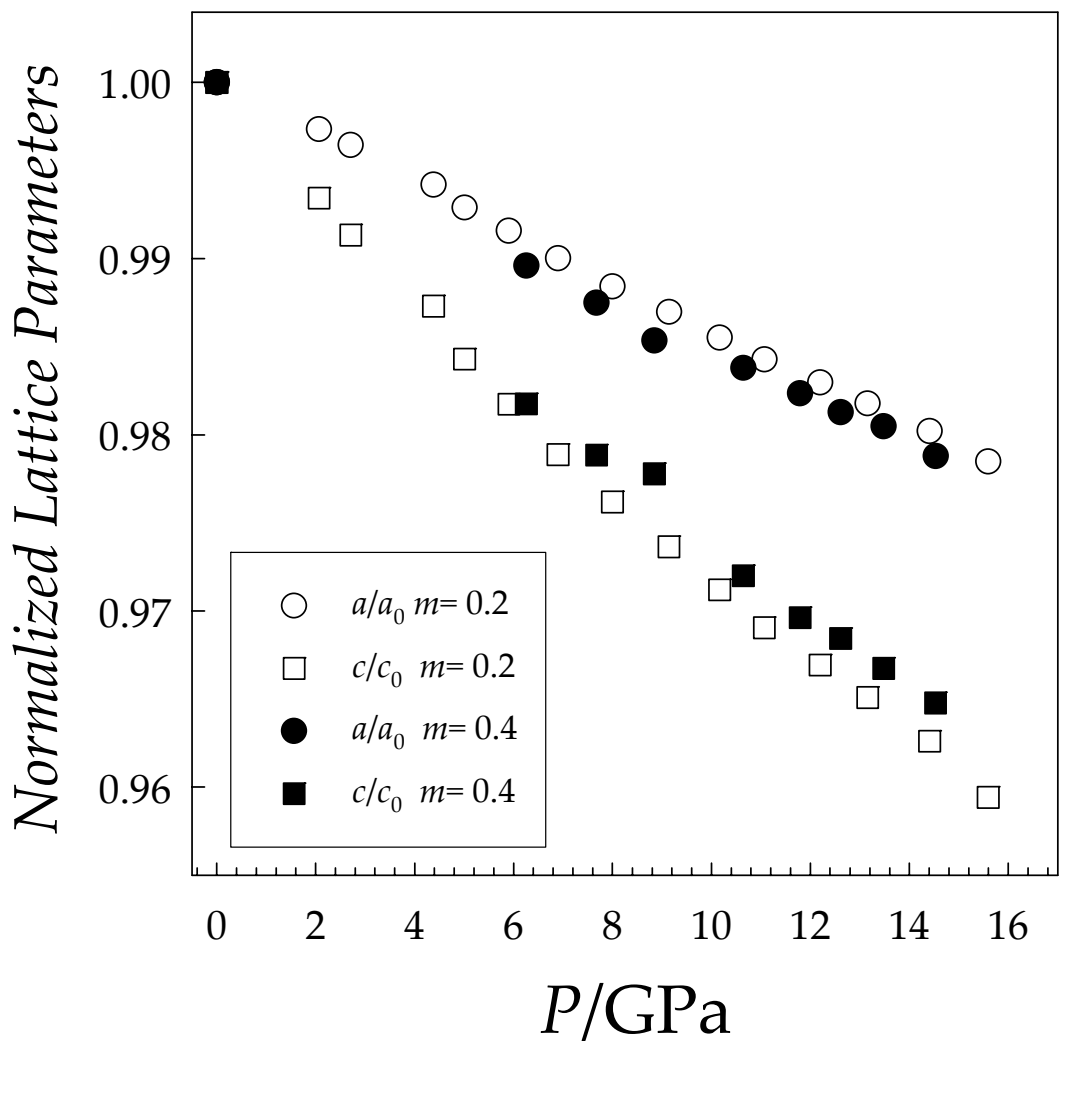

Figure 2

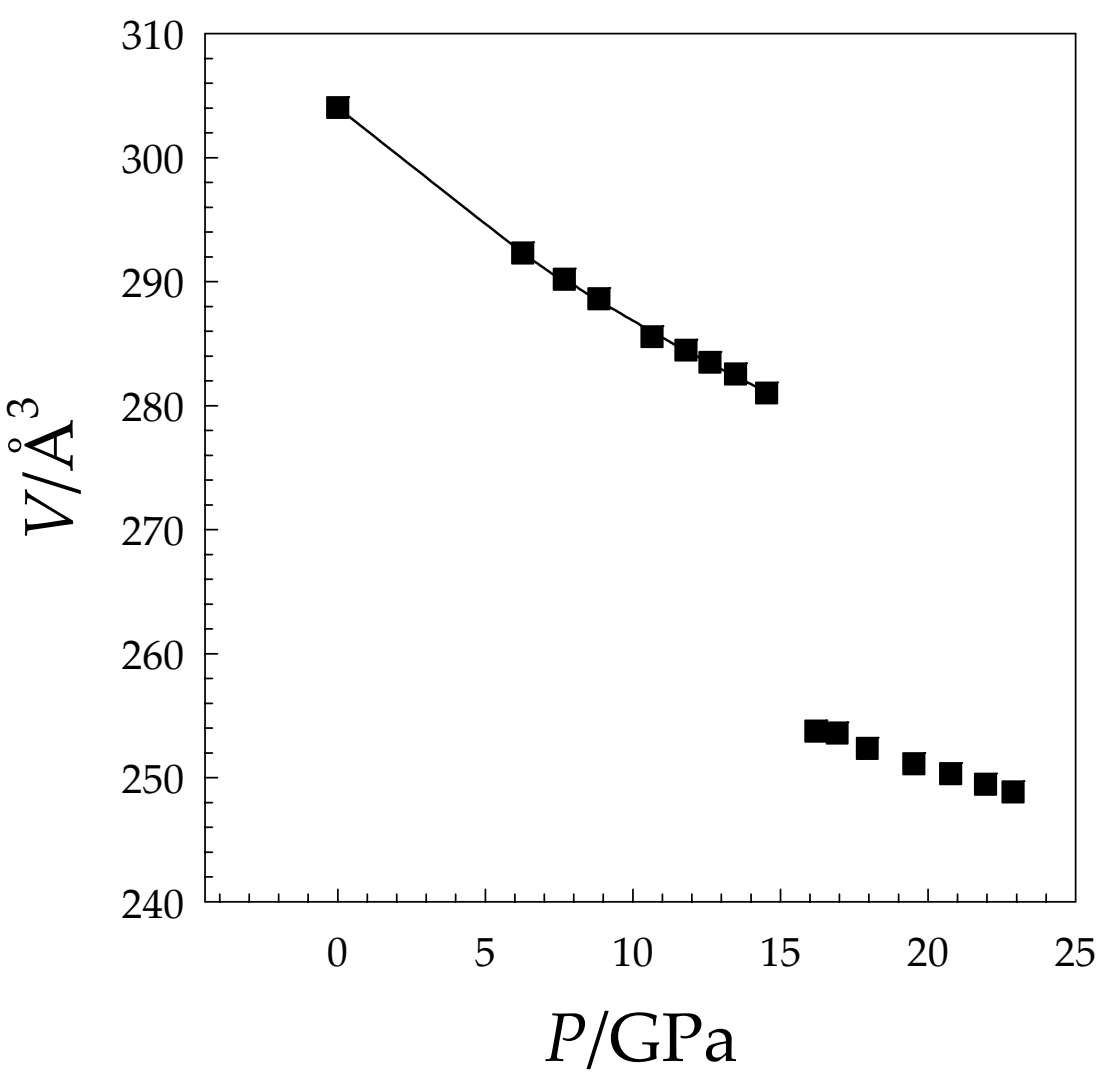

Figure 3

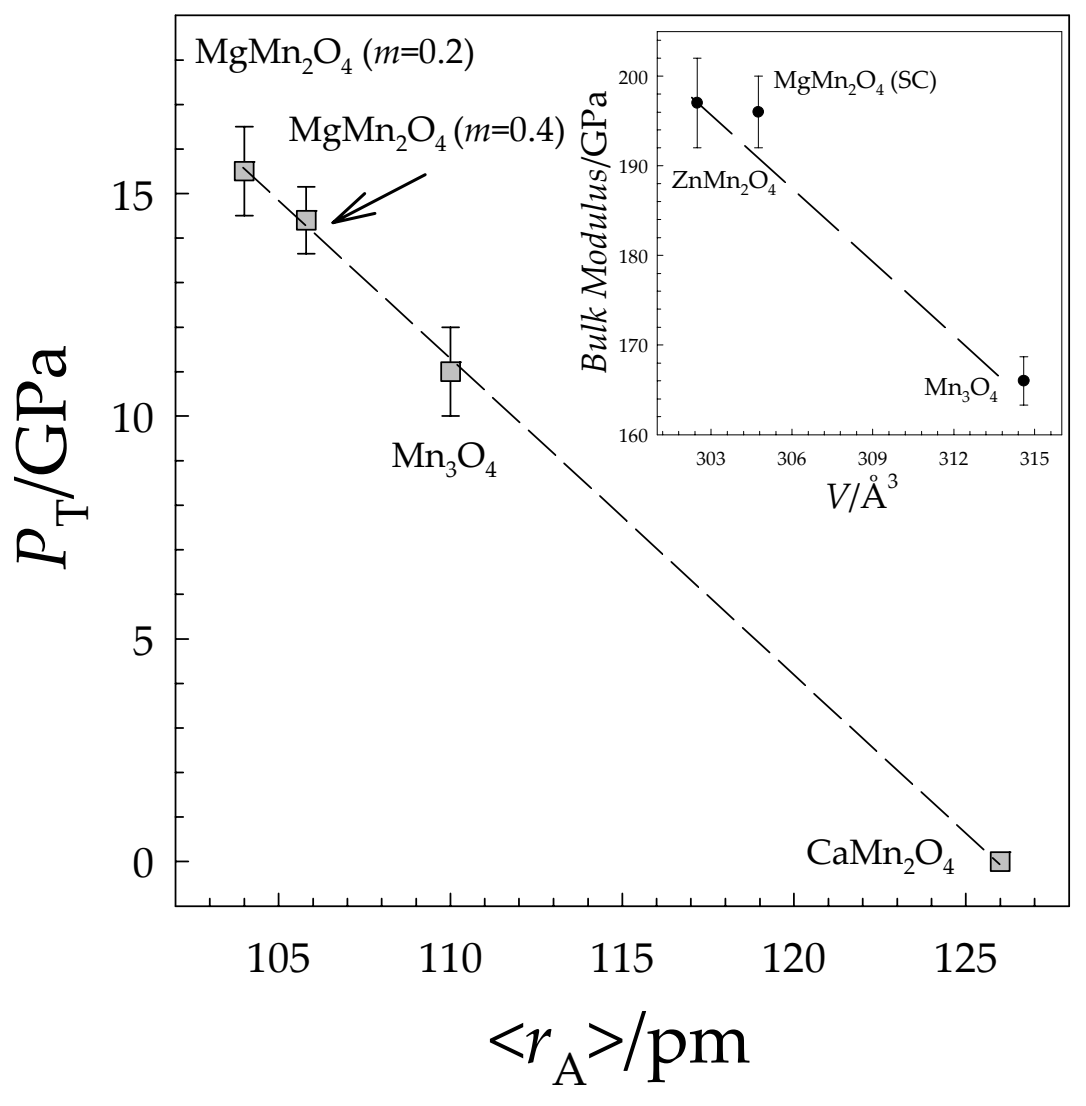

Figure 4